# Synthetic data: How could it be used for infectious disease research?

Styliani-Christina Fragkouli [1,2], Dhwani Solanki [3], Leyla J Castro [3], Fotis E Psomopoulos [2], Núria Queralt-Rosinach [4], Davide Cirillo [5], Lisa C Crossman [6,7]

[1]Department of Biology, National and Kapodistrian University of Athens, Athens, Greece
[2]Institute of Applied Biosciences, Centre for Research and Technology Hellas, Thessaloniki, Greece
[3]ZB MED Information Centre for Life Sciences, Cologne, Germany
[4]Department of Human Genetics, Leiden University Medical Center, Leiden, The Netherlands,
[5]Barcelona Supercomputing Center (BSC), Barcelona, Spain
[6]SequenceAnalysis.co.uk, Norwich Research Park, Norwich, UK
[7]School of Biological Sciences, University of East Anglia, Norwich Research Park, Norwich, UK

1. Introduction

Over the last three to five years, it has become possible to generate machine learning synthetic data for healthcare-related uses. However, concerns have been raised about potential negative factors associated with the possibilities of artificial dataset generation. These include the potential misuse of generative artificial intelligence (AI) in fields such as cybercrime, the use of deepfakes and fake news to deceive or manipulate, and displacement of human jobs across various market sectors.

Here we consider both current and future positive advances and possibilities with synthetic datasets. Synthetic data offers significant benefits, particularly in data privacy, research, in balancing datasets and reducing bias in machine learning models. Generative AI is an artificial intelligence genre capable of creating text, images, video or other data using generative models. The recent explosion of interest in GenAI was heralded by the invention and speedy move to use of large language models (LLM). These computational models are able to achieve general-purpose language generation and other natural language processing tasks. LLM are based on transformer architectures which improved on previous neural network architectures and were put forward by Vaswani et al. [1] where GenAI made an evolutionary leap from recurrent neural networks.

Fuelled by the advent of improved GenAI techniques and wide scale usage, this is surely the time to consider how synthetic data can be used to advance infectious disease research. In this commentary we aim to create an overview of the current and future position of synthetic data in infectious disease research [2].

## 2. Definition of synthetic data

The study of synthetic data (SD) has become a cornerstone in life sciences, with a growing interest both from academia and industry such as pharmaceutical companies. The emerging applications of SD include data augmentation, machine learning (ML) model training and validation, enhancing the statistical power of studies as well as supporting the development of new algorithms while ensuring the improvement of their fairness, bias and robustness [3].

A common way of describing SD is to consider it an artificially generated product from a model that attempts to capture the underlying distribution and structure of some real data. Despite its popularity no broad definition has been accepted by the scientific community so far and it usually comes down to the lens through which SD is viewed depending on the application or use case. However, D'Amico et al. [3] tries to fill this gap by proposing a definition that states "Synthetic data is data that has been generated using a purpose-built mathematical model or algorithm, with the aim of solving a (set of) data science task(s)".

Regarding clinical settings, the literature shows an increasing trend in SD implementation. There are many and varied types of SD ranging from tabular data, time-series, multi-omics, images, audio etc. [4]

## 3. Methods for synthetic data generation

Current SD generators include methods based on algorithms like Generative Adversarial Networks (GANs) and Variational Auto-encoders (VAEs). In addition, novel software has been developed by industry (e.g. Synthea, MDClone's Synthetic Data Engine [5]). Generally, SD can be divided into two groups, fully synthetic and partially synthetic, that is a blend of real and synthetic. VAEs and GANs are utilized in the generation of synthetic data that closely mimics real-world data. VAEs are deployed for the purpose of deriving latent representations from the input data, facilitating the creation of novel samples with comparable attributes. GANs have a significant impact on the generation of data through the training of a generator network that produces synthetic samples closely resembling original data, while a discriminator network differentiates between real and synthetic data. New to the table, large language models utilising transformer architecture have heralded a new era for generative AI.

## 4. Metadata, sharing and publishing synthetic data

Metadata provides a description of the information contained in a dataset and plays a crucial role in facilitating data sharing and integration. Metadata could include generic elements such as name, authors, and publication date of a dataset that help uniquely identify and clearly describe a dataset, facilitating its future retrieval and reuse [6]. It could also provide more specific elements such as how many data points, features, or target variables (e.g., for an ML classification prediction task). In areas related to (bio)medical research, the metadata provides a way to share enough information on the real data whilst also maintaining data privacy. Metadata concerning research objects, including datasets, software and more, are gaining recognition since it is easier to connect objects and thus get a more comprehensive picture of a research project.

For the particular case of ML, local, regional and international efforts are working on metadata schemas and guidelines. For instance, the German-based NFDI4DataScience project uses metadata to provide a harmonized view of ML-based research (datasets, software, ML models, tutorials and so). The Machine Learning Focus Group, part of ELIXIR Europe, established the Data, Optimization, Model and Evaluation (DOME) recommendations for publishing and sharing supervised ML approaches in biology aiming to improve evaluation and validation [7]. The ML Commons consortium recently released Croissant ML; a metadata schema based on schema.org for ML-ready datasets. Focused on ML models rather than datasets used for ML, the Research Data Alliance FAIR4ML Interest Group, including members of NFDI4DataScience, ELIXIR Europe, and EOSC, are also working on a schema.org extension.

Particularly for the case of SD metadata, ELIXIR Europe and NFDI4DataScience collaborated on an effort related to ML generation and usage [8]. This effort proposed a minimal metadata model based on schema.org and discusses the need of properties to characterize variables or features included in the dataset together with the number of features and the sample size. Some of these elements are also identified in Croissant ML as needed for ML-ready datasets in general.

## 5. Use in training machine learning models

Synthetic data has been successfully used to train AI models without the use of any real-world data. The most important reasons for the use of synthetic data in model training include privacy preservation and bias mitigation (see below) and for use in balancing datasets. Healthcare providers in fields such as medical imaging use synthetic data to train AI models while protecting patient privacy. For example, startup Curai trained a diagnostic model on 400,000 simulated medical cases.

In machine learning, a dataset used for classification with skewed class proportions is termed imbalanced. Classes making up the largest proportion of the dataset are known as the majority classes with those making up the smaller proportion known as minority classes. Biological data is characterized by being noisy, with a high amount of variation, dimensionality and complexity. Several years of pathogen research have indicated that high consequence pathogens in an outbreak are not only sequenced and studied with higher frequency than environmental organisms but are mainly clonal with small amounts of variation (as seen in SARS-CoV-2 with COG UK, COVID-19 Genomics COnsortium). When studying such data and populations for traits using machine learning, care should be taken to balance classes across the dataset. Given sufficient balanced samples, this may be relatively simple, however, other methods may be required in some cases. Traditional strategies for balancing classes can include downsampling and/or upsampling to create a more even dataset, synthetic data could also be of use in this regard.

In the context of imbalanced data generation *via* Generative Adversarial Networks (GANs), maintaining the correlation and dependencies among categorical variables poses a significant challenge. This issue represents a critical limitation of imbalanced data. In addition, the predominant class often exerts significant influence during training, leading to challenges for the discriminator in recognizing the underrepresented class and impeding the generator's ability to accurately represent minor classes. To avoid these challenges, a differentially private conditional Generative Adversarial Network model could be the appropriate method that mainly focuses on data transformation, sampling, conditioning, and network training to generate realistic and privacy-preserving personal data. It also handles the mixed data types and correlations between variables [9]. In the most recent LLM models, bias and hallucinations may be introduced into the dataset due to synthetic data generated from LLMs in the training phase. The incorporation of regularization methods is essential to enhance the stability of training when dealing with datasets that contain noise [10].

## 6. Examples of synthetic data use in infectious disease research

### 6.1 CT scans and chest X-rays for use in COVID-19 diagnostics

A fully synthetic dataset was built for the diagnosis of COVID-19 by Zunair & Hamza [11]. The authors of the study noted a performance improvement in detecting COVID-19 in patients when using synthetic images. The use of CT scan and X rays can aid in determination of which patients are at higher risk of severe illness and may require intubation. At the time of publication, there was only one suitable COVID-19 X-ray image dataset containing positive or suspected cases as well as other viral or bacterial pneumonia cases. Negative cases were acquired from publicly available datasets and merged, to create a training dataset which was somewhat imbalanced and created poor performance on detecting the minority class – the vital positive for COVID-19. The authors created a synthetic dataset by oversampling positive cases after carefully investigating saliency maps to define activations using a heatmap approach to uncover the most important features the network was using for discriminatory analysis.

An alternative synthetic dataset was formed by Karbhari *et al*. [12] in which GAN were employed to generate a synthetic dataset of 500 COVID-19 radiographic images.

## 6.2 Synthetic data for benchmarking wastewater surveillance pipelines

Sutcliffe *et al* (2024) examined the use of wastewater-based surveillance (WBS) to monitor pathogens, particularly SARS-CoV-2, emphasising its role during the COVID-19 pandemic [13]. Historically, WBS has been critical for tracking poliovirus and has evolved to estimate community infection levels and genome sequencing, revealing the genetic diversity of viruses in wastewater due to a mix of viral lineages and sequencing errors that is present in wastewater samples making the identification process challenging. Nine computational tools, designed to differentiate SARS-CoV-2 lineages in simulated wastewater samples, were evaluated, including Delta and Omicron variants as well as synthetic 'novel' lineages. The tools generally performed well in identifying lineages and estimating their relative abundances, even in low-frequency lineages (down to 1%), although accuracy improved with a 5% frequency threshold. The presence of unknown synthetic lineages slightly increased error in abundance estimates, and the tools varied in handling novel or recombinant lineages. The study highlights the need for continued development and benchmarking of computational tools through the implementation of synthetic data to address the evolving challenges in detecting and quantifying viral lineages from complex wastewater samples, which is crucial for effective public health surveillance.

Another distinct activity of synthetic data in wastewater surveillance is in response to a significant challenge: determining the level of detection confidence based on the frequency and coverage of microbial presence. As such, allowing users and analysts to explore scenarios of low-input microbial presence while evaluating classification accuracy can directly address this challenge. An example workflow, that has been implemented as part of an ELIXIR BioHackathon project (https://biohackathon-europe.org) includes taxonomic classification, detection of Single Nucleotide Variants (SNVs), and identification of novel taxa, variants, and lineages, with the primary outcome being a quantifiable measure of ground-truth as a probability.

## 6.3 Epidemiology and Pandemic modelling

The COVID-19 pandemic has emphasized the critical need for accessible data in public health surveillance, clinical research, and policy analysis. Access to data is essential for developing and applying mathematical and computational methods to model and analyse the spread and impact of infectious diseases, thereby enabling effective and timely management of disease outbreaks. These methods typically include equation-based Compartment Models (CMs), which simulate population-level dynamics of disease transmission and progression, and Agent-based Models (ABMs), which simulate the behaviour of individual agents within a population. In recent years, particularly during the COVID-19 pandemic, machine learning (ML) and artificial intelligence (AI) have emerged as complementary approaches in disease modelling, especially used to uncover patterns in large datasets and predict disease outcomes.

Due to the critical importance of data access and modelling quality, synthetic data is increasingly employed in this context. Synthetic data is proven to accelerate research, support real-time epidemiological tracking, enable comprehensive sensitivity analyses, and enhance the accuracy of predictive algorithms [14]. Various types of synthetic data have been created using diverse generative approaches (statistical methods, machine learning, simulations), tailored for different disease modelling purposes as described above. Moreover, hybrid approaches using both real and synthetic data have been developed. Examples include using high-quality conditional synthetic data alongside real data to improve machine learning models in context with limited pandemic data and label scarcity [15] and using short-term predictions derived from real data curve fitting as synthetic data to enhance long-term forecasts during the COVID-19 pandemic [16].

The use of synthetic data not only enables improving performances of pandemic modelling but opens avenues for new insights in infectious disease research. For example, Zhu *et al.* introduced a methodology to create synthetic populations specifically for infectious disease transmission simulations, highlighting the significance of household structure relationships [17]. Meanwhile, Popper *et al.* utilized an agent-based simulation model to generate synthetic COVID-19 case data in Austria, offering researchers a more detailed dataset, including unobservable details such as the moment of infection [18].

Several large-scale projects have produced high-fidelity synthetic datasets now available for research. For instance, the U.S. National COVID Cohort Collaborative (N3C) synthetic dataset accurately mirrors real COVID-19 patient data, replicating results obtained from analyses conducted on actual patient records. Another example is the high-fidelity synthetic COVID-19 dataset within the UK Clinical Practice Research Datalink (CPRD) Aurum database. The dataset includes comprehensive data on sociodemographic factors and clinical risk profiles, covering patients with COVID-19 symptoms and controls testing negative for COVID-19 in primary care settings.

## 6.4 Digital Twins

The digital twin is a full virtual synthetic representation of an object or system designed to accurately reflect a physical object. The twin spans the object's lifecycle, is updated from real time data and employs simulations, machine learning and reasoning to aid in the making of decisions. The core idea of having a digital twin comes from manufacturing processes and was essentially pioneered by NASA in the 1960s. Various areas of science, including but not limited to engineering, climate science, agriculture and social science have realised the potential of digital twins. Medical digital twins combining an understanding of physiology and viral replication with AI models from population and clinical data are promising as tools to optimize treatment of patients infected with a virus [19]. Some recent examples of digital twin use in healthcare include oncology, artificial pancreas, heart patients, migraine care, drug discovery and clinical trials. One area a validated digital twin could advance is to greatly reduce the cost and complexity of studying effectiveness of combinations of different clinical interventions. However, to properly personalise a digital twin, a model would need to integrate with clinical records and time courses such as vital signs, immune cell counts and CT scans of infected organs as well as treatment responses. Since the immune system plays an important role in a wide range of diseases and health conditions, from autoimmune disorders to fighting pathogens, digital twins of the immune systems will have an especially high impact. Whilst still in its infancy for use in infection, community efforts such as COMBINE (Covid-19 disease map project and computational modelling in Biology network) are working towards digital twin technology. Big Pharma company GSK together with Siemens and Atos have begun implementing digital twins into development of vaccines.

## 7. Challenges

A suite of challenges exist to incorporate synthetic data in infectious disease research, these include privacy, bias, technical and economic concerns as well as trust and acceptance.

**7.1 Data privacy** is defined as the ability of a person to determine for themselves, when, how and to what extent personal information is shared with or communicated to others. These rights are underlined in the most comprehensive privacy regulation worldwide, European general data protection regulation (GDPR), however, legislation differs according to region. Maintaining good security of the systems storing the real data can require significant technical safeguards.

**7.2 Bias** refers to a systematic discrepancy or persistent deviation that arises during the data sampling or testing process. This can lead to the overestimation or underestimation of risks related to certain clinical outcomes. Such bias can stem from the population being studied, the techniques used to collect data, or the methods employed to derive the new dataset. Equitable AI is a field to balance these challenges and ensure equal representation of data. Clearly, apart from the diversity and representativeness of data, accuracy is also important. It is known that genetic databases in biobanks lack diversity, however, new approaches to international strain sequencing of SARS-CoV-2 indicate that more international collaboration is within reach.

**7.3 Technical and economic** challenges exist where models have some limitations in terms of use. Feature engineering, or the methods used to encode the data input to the model can be crucial and often requires a deep understanding of the complexity and data type with specialist knowledge. Resource and economic management are required to determine the cost effectiveness of model use, carefully balanced with their accuracy. Potential stumbling blocks also include integration with current workflows, updating and security of the models, ownership and legal ramifications in case of the provision of incorrect data.

**7.4** Synthetic data is affected by concerns of **trust and acceptance**. To make more use of this data type, people need to trust in the data authenticity and reliability. To increase trust and acceptance, robust methods to validate the fidelity of data and utility are required. Explainable AI (XAI) techniques are essential for making AI methods interpretable and transparent, thus shedding light on the black-box effect, especially when working with synthetic data. XAI methods allow users to comprehend the underlying mechanisms and how AI models reach their decisions, by providing insights into the relationship of input and output and whether biases are present. It should be noted that there are some challenges in these methods too, for instance they can prove subjective and context dependent [20].

## 8. Concluding remarks

Data is a foundation of the modern economy and may be 'the world's most valuable resource' [the Economist]. It has been estimated that by 2030, data obtained from direct measurements will be constrained by cost, logistics and privacy reasons, meaning that synthetic, artificially created data from simple rules, statistical models and AI will completely overshadow real data in AI models [Gartner].
 Overall, despite concerns, synthetic data can have many uses in infectious disease research. Data privacy issues relating to human genome sequences require technical safeguards which include encryption and cryptographic tools, k-anonymity (anonymised datasets) and specialist federated machine learning (differential privacy). A further safeguard is synthetic data generation. Whilst privacy of direct human whole genome sequences is of the most concern, privacy in metagenomic data is also important. It is known that metagenomic data, particularly from infectious disease clinical sources can also contain human sequences; other sources of metagenomic data requiring privacy include sequences and resources covered under the Nagoya Protocol. Synthetic data and the digital twin have the potential to create a revolution in drug development, including the creation of synthetic patient platforms to allow assessing new molecules under different patient scenarios to allow discovery speedup.

A careful understanding, scrutiny and quality control is needed of the data and any associated platforms to ensure that biases, representativeness and data accuracy are strictly considered. One important dynamic to consider is the trade-off between fidelity (the degree of similarity to real data) and the synthetic (or the sim-to-real gap), particularly important to patient privacy. Through the careful incorporation of synthetic data to emulate all types of conditions and individuals, AI can better make decisions for our healthy futures.